\documentclass[prb,onecolumn,superscriptaddress,floatfix,showpacs,amsmath,amssymb]{revtex4-2}
\usepackage{graphicx}% Include figure files
\usepackage{dcolumn,xcolor}% Align table columns on decimal point
\usepackage{bm}% bold math
\usepackage{color}
\usepackage{tabularx}
\usepackage{float}
\usepackage{siunitx}
\usepackage{verbatim}
\usepackage{parskip}
\usepackage{soul,xr}
\usepackage[normalem]{ulem}
\newcommand{\Ahmed}[1]{\textcolor{black}{#1}}

\newcommand{\REZO}{$RE_2$Zr$_2$O$_7$}
\newcommand{\CZO}{Ce$_2$Zr$_2$O$_7$}
\newcommand{\PZO}{Pr$_2$Zr$_2$O$_7$}
\newcommand{\NZO}{Nd$_2$Zr$_2$O$_7$}

\newcommand{\TZO}{Tb$_2$Zr$_2$O$_7$}
\newcommand{\DZO}{Dy$_2$Zr$_2$O$_7$}
\newcommand{\HZO}{Ho$_2$Zr$_2$O$_7$}

\newcommand{\DTO}{Dy$_2$Ti$_2$O$_7$}
\newcommand{\HSO}{Ho$_2$Sn$_2$O$_7$}
\newcommand{\HTO}{Ho$_2$Ti$_2$O$_7$}
\newcommand{\TTO}{Tb$_2$Ti$_2$O$_7$}

\newcommand{\hoion}{Ho$^\mathrm{3+}$}

\newcommand{\mb}{$\mu_{\rm B}$}

\newcommand{\tb}{$T_{\mathrm b}$}

\newcommand{\chip}{$\chi_{\mathrm {ac}}^{\prime}$}
\newcommand{\chipp}{$\chi_{\mathrm {ac}}^{\prime\prime}$}
\newcommand{\mbho}{$\mu_{\mathrm {B}}/{\rm Ho^{3+}}$}

\usepackage{xr}
\makeatletter
\newcommand*{\addFileDependency}[1]{
  \typeout{(#1)}
  \@addtofilelist{#1}
  \IfFileExists{#1}{}{\typeout{No file #1.}}
}
\makeatother

\newcommand*{\myexternaldocument}[1]{
    \externaldocument{#1}
    \addFileDependency{#1.tex}
    \addFileDependency{#1.aux}
}
\myexternaldocument{HOZO_SI}

\begin{document}

\title{Slow Spin Relaxation and Low-Temperature Spin Freezing in \Ahmed{Disordered Fluorite} Ho$_2$Zr$_2$O$_7$}
	%%% Authors%%% Slow spin relaxation and low-temperature spin freezing in disordered fluorite Ho2Zr2O7
	\author{A.~Elghandour}
	\email{ahmed.elghandour@kip.uni-heidelberg.de}
	\affiliation{Kirchhoff Institute for Physics, Heidelberg University, INF 227, 69120 Heidelberg, Germany}

        \author{Sheetal}
	\affiliation{School of Physical Sciences, Indian Institute of Technology Mandi, Mandi-175075 (H.P.), India }\altaffiliation{Current affiliation: Jülich Centre for Neutron Science (JCNS-4) at Heinz Maier-Leibnitz-Zentrum (MLZ), Forschungszentrum Jülich GmbH, Lichtenbergstr. 1, 85747 Garching, Germany}

        \author{C.~S.~Yadav}
	\affiliation{School of Physical Sciences, Indian Institute of Technology Mandi, Mandi-175075 (H.P.), India }
 	
  \author{R.~Klingeler}\email{klingeler@kip.uni-heidelberg.de}
  \affiliation{Kirchhoff Institute for Physics, Heidelberg University, INF 227, 69120 Heidelberg, Germany}

\date{\today}
\begin{abstract}
We report on the origin of spin freezing in \Ahmed{the disordered fluorite \HZO. The system is investigated} by low-temperature heat capacity as well as by DC and AC magnetization. While the system does not show a long-range magnetic order down to at least $T = 280$~mK, we observe signatures of slow spin dynamics, magnetic field induced relaxation processes at relatively high temperatures, and a spin-frozen state below $T = 0.6$~K. Our results suggest that similar to the canonical spin ice systems \HTO~\cite{ehlers2004evidence} and \DTO~\cite{snyder2001spin}, spin freezing in \HZO\ is preceded by two slow spin relaxation processes; the first forms a field-induced region extending to at least 18~K and the second is rather field-independent and appears at $T_{\mathrm g2} = 1$~K~\cite{ramon2020geometrically}.

\end{abstract}

%\pacs{}

\maketitle

\section{Introduction}

Antiferromagnetically coupled spin on geometrically frustrated lattices such as pyrochlore or triangular can preserve strong fluctuations and prevents the establishment of long-range order even at $T \rightarrow 0$. The unveiled exotic phases are characterized by intriguing properties such as spin ice, spin liquid, spin glass, plateau phases and fractionalized spin excitations~\cite{harris1997geometrical,Savary2017,Zhoureview,Broholm2020,Ning2023}. The pyrochlore oxides \HTO\ and \DTO\ have garnered significant research interest for almost two decades due to their distinctive property of hosting monopole-like excitations~\cite{harris1997geometrical,ramirez1999zero}. The variations in the rare-earth site modulates the magnetic ground state from classical spin ice exemplified by \HTO\ and \DTO\ to quantum spin ice observed in \CZO\ and \NZO~\cite{den2000dipolar,gao2019experimental,gaudet2019quantum,gao2022magnetic,blote1969heat,lhotel2015fluctuations,xu2015magnetic}.

The zirconate compounds of \REZO -type where RE = Ce, Pr, Nd crystallize in an ordered fluorite (pyrochlore) structure with space group $Fd\overline{3}m$ and RE = Tb, Dy, Ho crystallizes in a defect-fluorite/pyrochlore structure~\cite{reynolds2013anion}. Introducing the structural disorder is found to significantly alter the physical properties; for instance, a structurally clean \DTO\ pyrochlore hosts spin ice ground state, while \DZO\ adopts a chemically disordered structural phase, exhibiting a dynamic ground state down to 50~mK without zero-point entropy~\cite{ramirez1999zero,ramon2019absence}. A similar scenario is observed in \HTO , where chemical alterations within the pyrochlore phase hinder the emergence of spin-ice correlations in \HZO~\cite{harris1997geometrical,Sheetal_2022}. \Ahmed{The pyrochlore system \TTO\ retains a dynamic magnetic ground state, while even a subtle lattice disorder induces a spin-glass behavior in Tb$_2$Hf$_2$O$_7$ which is an exemplary case of the anion-disordered pyrochlore~\cite{Sibille2017}. This contrast underscores the broader implications of structural disorder on the physical properties of pyrochlore systems.}

In this paper, we investigate the magnetic ground state of \HZO, employing both DC and AC magnetization measurements down to low temperatures of 400~mK. In doing so, we found that unlike \DZO, the introduction of a chemical disorder does not induce spin dynamics but rather stabilizes the spin freezing at $T < 0.6$~K.  Moreover, we examine the spin dynamics and magnetic field-induced relaxation processes at relatively high temperatures, serving as a precursor phase to the spin freezing observed at $T < 0.6$~K.  Our results suggest that similar to the canonical spin ice systems \HTO~\cite{ehlers2004evidence} and \DTO~\cite{snyder2001spin}, the spin freezing in \HZO\ is preceded by two slow spin relaxation processes; the first process forms a field-induced region which extends to $T \simeq 18$~K and the second is rather a field-independent and it appears at $T_{\mathrm g2} = 1.1 (1)$~K~\cite{ramon2020geometrically}.

\section{Experimental Methods}

Phase pure polycrystalline \HZO\ has been prepared by the solid-state reaction method as reported in Ref.~\cite{Sheetal_2022}. Our studies were performed on thin sliced pellets with $m=7.8(2)$~mg. DC magnetization was measured in the temperature range of 1.8 to 60~K and in static magnetic fields up to 4~T by means of the Magnetic Properties Measurement System (MPMS3, Quantum Design). For measurements down to 400~mK, the MPMS3 was equipped with the iQuantum $^3$He probe. Both field-cooled (fc) measurements, where the sample was cooled in the measurement field, and zero-field-cooled (zfc) measurements were performed, where the sample was cooled to the lowest temperature before applying the external magnetic field. Prior to each isothermal DC magnetization measurement using $^3$He probe, the sample's temperature was elevated to 10~K in the absence of a magnetic field, then cooled to the target temperature to start the measurement. AC magnetization was measured in the temperature range of 1.8 to 60~K, with a $B_{\rm ac}$~=~7 - 9~Oe AC excitation field, up to 4~T DC magnetic fields, and frequencies ranging from 3~Hz to 400~Hz using the AC option of the MPMS3. For measurements at $f>1$~kHz, the ACSMII option of the Physical Properties Measurement System (PPMS, Quantum Design) was employed.

\section{Experimental Results}
\subsection{Spin dynamics at high temperatures}

\begin{figure}
    \centering
    \includegraphics[width=1\columnwidth,clip]{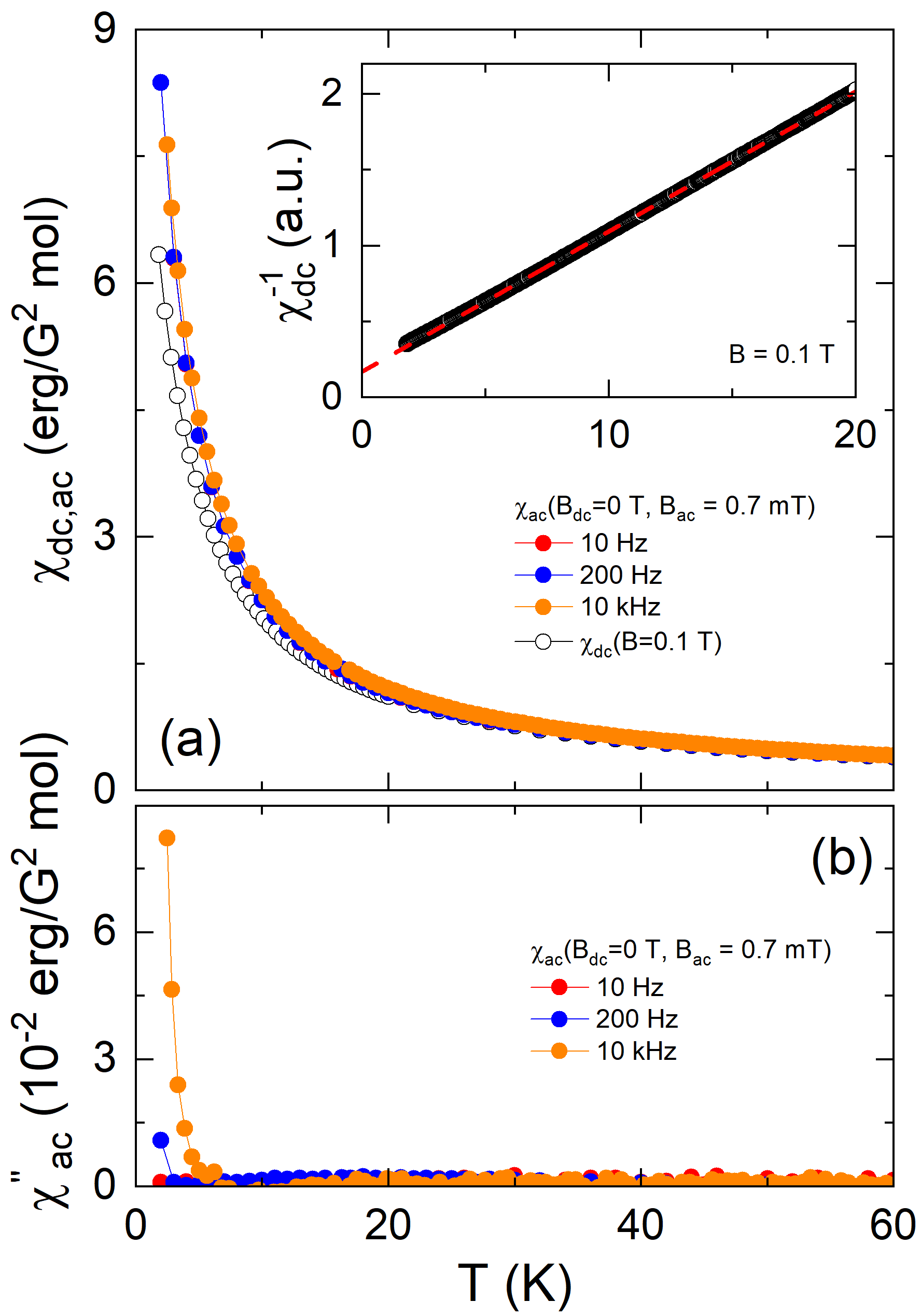}
    \caption{Temperature dependence of (a) real part $\chi'$ and (b)  imaginary part $\chi''$ of AC magnetic susceptibility measured at different frequencies at $B_{\rm dc} = 0$~T ($B_{\rm ac} = 0.7$~mT). Black open circles in (a) show the static magnetic susceptibility $\chi_{\rm dc}=M/B$ measured at $B=0.1$~T. Inset: Inverse of $\chi_{\rm dc}$ and Curie-Weiss fit to the data (dashed red line).
}
    \label{ac_zerofield}
\end{figure}

Figure~\ref{ac_zerofield}a shows the temperature dependence of the real part, $\chi'_{\rm ac}$, of the AC magnetic susceptibility measured at different frequencies ($B_{\rm ac}=0.7$~mT) in zero DC magnetic field as well as the static magnetic susceptibility $\chi_{\rm dc}=M/B$ measured at $B=0.1$~T. Upon cooling, $\chi'_{\rm ac}$ increases monotonically following a Curie-Weiss-like behaviour. The data show no frequency dependence in the regime 10~Hz~$\leq f \leq$~10~kHz and $\chi'_{\rm ac}$ coincides with $\chi_{\rm dc}$. Similarly, $\chi''_{\rm ac}$ presented in Fig.~\ref{ac_zerofield}b does not show any anomalies either.

Curie-Weiss behaviour in $\chi_{\rm dc}$ is highlighted in the inset of Fig.~\ref{ac_zerofield}a. \Ahmed{Fitting the high-temperature data in various temperature regimes even down to 20~K}~\footnote{\Ahmed{Using different temperature regimes between 1.8~K and 300~K did not significantly change the results and was used to assess the uncertainties of the resulting parameters.}} by an extended Curie-Weiss law $\chi_{\rm dc}=C/(T+\Theta)+\chi_0$ where $\chi_0$ is a temperature-independent term, yields the Weiss temperature $\Theta = -3(1)$~K. From the obtained Curie constant $C$ we deduce the effective magnetic moment $\mu_{\rm eff}\simeq 9.5(1)$~\mb /Ho$^{3+}$ in good agreement with Ref.~\cite{ramon2020geometrically} and with the expected free ion value.

The sign of the estimated Weiss temperature and its small value indicate predominance of weak antiferromagnetic couplings of a few Kelvin, which agrees well with previous reports~\cite{ramon2020geometrically}. Using a mean-field approach and assuming only nearest-neighbour (nn) couplings, the average magnetic interaction $J_{\rm nn}$~\footnote{$J_{\rm nn}$ is defined using ${\cal H}=J_{\rm nn}\sum_{ij}S_iS_j$.} between \hoion\ moments can be estimated by~\cite{stanley1987introduction}

\begin{equation}
 J_{\rm nn} =\frac{3\Theta}{zJ(J+1)}.
    \label{exchange interaction}
\end{equation}

Assuming the number of nearest neighbors is $z=6$ and the total angular momentum $J=8$ yields $J_{nn} = -20(6)$~mK which is in a good agreement with previous reports~\cite{ramon2020geometrically}.

\begin{figure}
    \centering
    \includegraphics[width=1\columnwidth,clip]{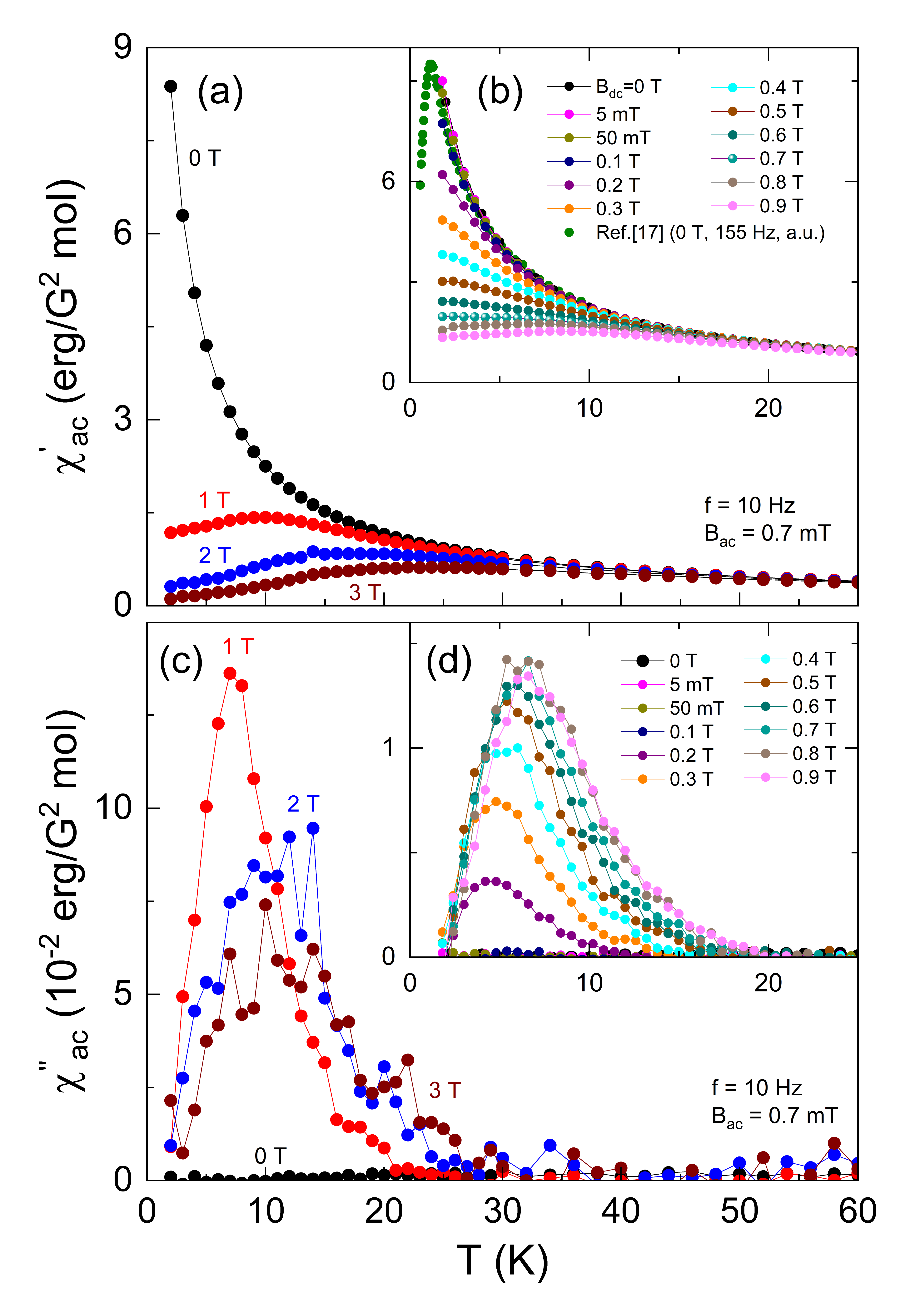}
    \caption{Temperature dependence of (a,b) the real part \chip\ and (c,d) the imaginary part \chipp\ of the AC magnetic susceptibility of \HZO\ measured at different static magnetic fields 0~T~$\leq B \leq $3~T, 0.7~mT AC excitation field, and $f=10$~Hz. (b,d) show the data measured at $B < $~1~T. In (b) also data measured at $B=0$~T from Ref.~\cite{ramon2020geometrically} are shown which have been obtained at 155~Hz, $B_{\rm ac}$ unknown; these data have been normalised to our data.}
    \label{ac_infield}
\end{figure}

Applying external DC magnetic fields results in a more complex behavior of both $\chi'_{\rm ac}$ and $\chi''_{\rm ac}$ as shown in Fig.~\ref{ac_infield}. At $B_{\rm dc} > 0.1$~T, the AC susceptibility data measured as a function of temperature at $f=10$~Hz show significant dissipative effects below about 22~K in $\chi''_{\rm ac}$ and non-linear field dependence in $\chi'_{\rm ac}$ up to $\sim 30$~K. Note that in the same temperature regime, DC magnetization shows CW behaviour and very weak exchange interactions of $J_{\rm nn} \simeq 20(2)$~mK. The main features are: in $\chi'_{\rm ac}$, there is an anomaly which shifts to higher temperatures with applied field and develops into a broad hump at $B_{\rm dc} \ge 0.9 $~T. Further, $\chi'_{\rm ac}$ decreases upon cooling below this anomaly. In $\chi''_{\rm ac}$, a magnetic field of $B_{\rm dc} = 0.2 $~T induces a peak at $T = 4.2$~K. The peak's width and height are field dependent and its center shifts to higher temperatures with increasing the applied magnetic field. These features in $\chi'_{\rm ac}$ and $\chi''_{\rm ac}$ agree with the Kramer-Kronig relations.

\begin{figure}
    \centering
    \includegraphics[width=1\columnwidth,clip]{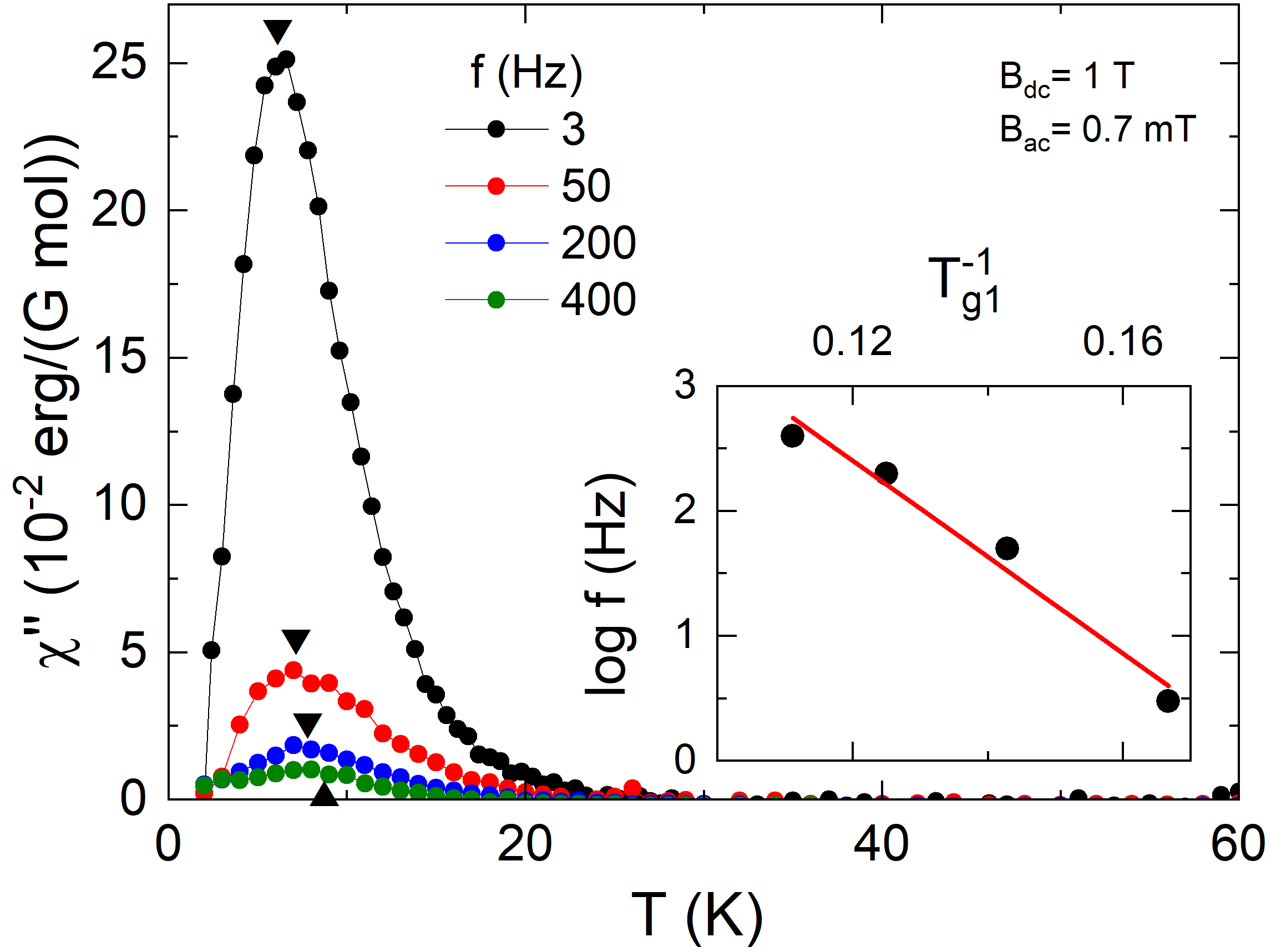}
    \caption{Temperature dependence of the imaginary part \chipp\ of the AC magnetic susceptibility of \HZO\ measured at different frequencies 3~Hz~$\leq f \leq $400~Hz, 0.7~mT AC excitation field, and $B_{\rm dc} = 1$~T. Black triangles mark the peak maxima $T_{\rm g1}$. Inset: Logarithm of measurement frequency versus the inverse of $T_{\rm g1}$. The red line is a fit by means of an Arrhenius law (see the text).}
    \label{ac_arrhenius}
\end{figure}

Figure \ref{ac_arrhenius} shows the temperature dependence of $\chi''_{\rm ac}$ measured at different frequencies with applied magnetic field of $B_{\rm dc} = 1$~T. The magnetic field induces a peak which shifts to higher temperatures with increasing the frequency. Additionally, the peak's width and height are frequency dependent. The data in Figs.~\ref{ac_infield} and~\ref{ac_arrhenius} hence imply a DC field-induced relaxation process in \HZO. Similar field-induced processes are typical features for the canonical spin ice systems \DTO~\cite{snyder2001spin} and \HTO~\cite{ehlers2004evidence} as evidenced by peaks in $\chi'_{\rm ac}$ and $\chi''_{\rm ac}$ of both systems at 15(1)~K. Likewise, other pyrochlore systems such as \TTO\ and \TZO\ show field-induced processes at 20~K and 25(1)~K, respectively~\cite{ueland2006slow,ramon2023glassy}.

\begin{figure}
\centering
\includegraphics[width=1\columnwidth,clip]{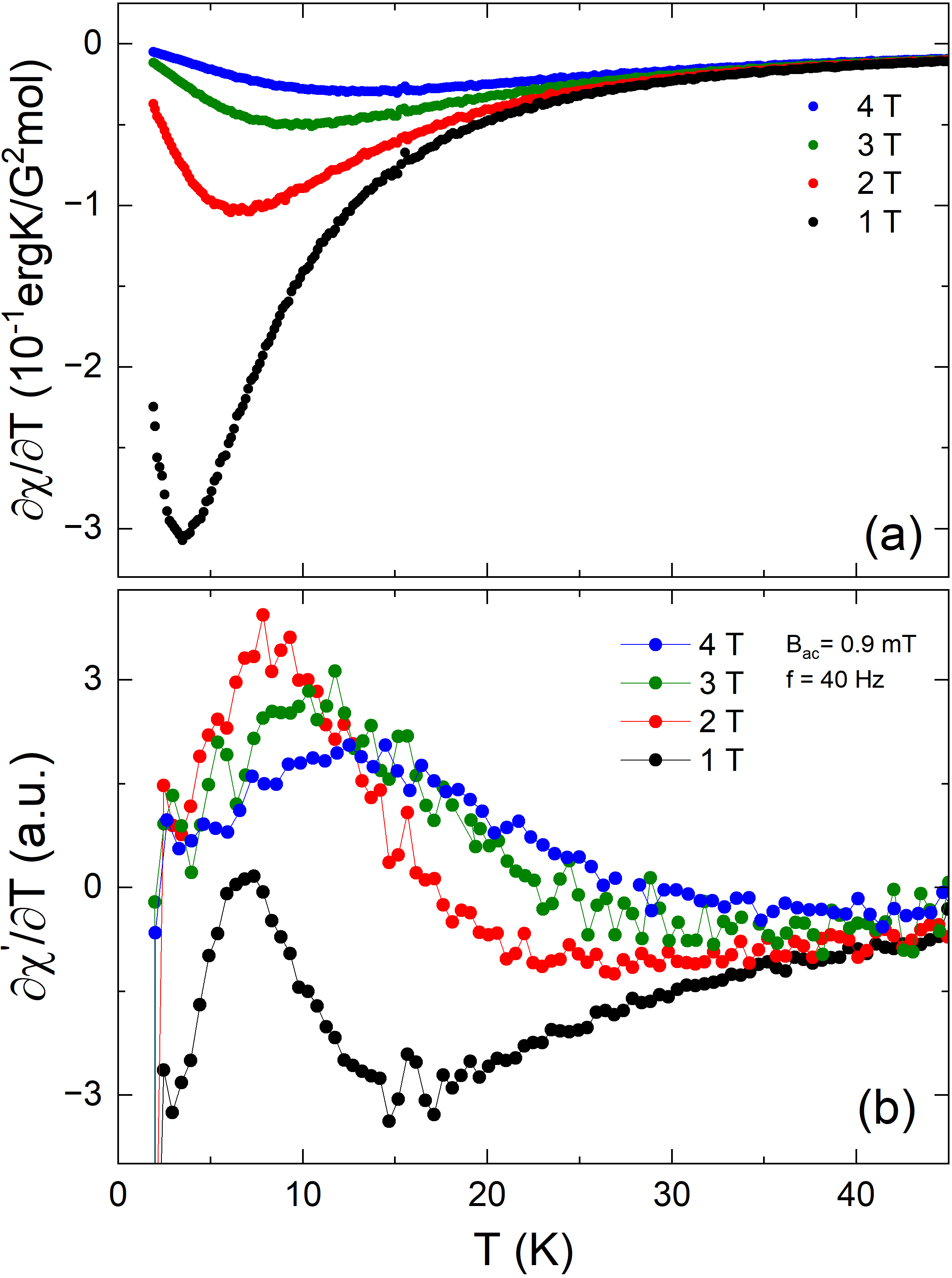}
\caption{Temperature dependence of the derivative of (a) the static magnetic susceptibility  $(\partial(M/B)/\partial T)$ measured at different static magnetic fields and (b) of the real part of the AC magnetic susceptibility $\partial\chi'/\partial T$, measured at the same applied DC fields.
}
    \label{derivative}
\end{figure}

The above mentioned field-induced anomalies in AC magnetisation also show up in the DC magnetisation. This is demonstrated by the temperature derivatives of $\chi_{\rm dc}$ and $\chi'$ (at $f=40$~Hz) displayed in Fig.~\ref{derivative}. The data show that the magnetic field induces clear features in the derivatives of both quantities which appear as minima and maxima, respectively. These features are gradually suppressed and shifted to higher temperatures upon increasing the field. Note, that the minimum in the DC magnetisation data appears at lower temperatures as compared to the $f=40$~Hz maxima in the AC magnetisation data. Further, the absolute magnitude of the peaks in $\partial(\chi')/\partial T$ is smaller than that in $\partial(\chi_{\rm dc})/\partial T$. A similar behaviour was observed in \TTO\ and has been attributed to the development of correlated clusters coupled through dipolar interactions~\cite{ueland2006slow}.

\subsection{Spin freezing below $T_{\rm b}=0.6$~K}

\begin{figure}
    \centering
    \includegraphics[width=1\columnwidth,clip]{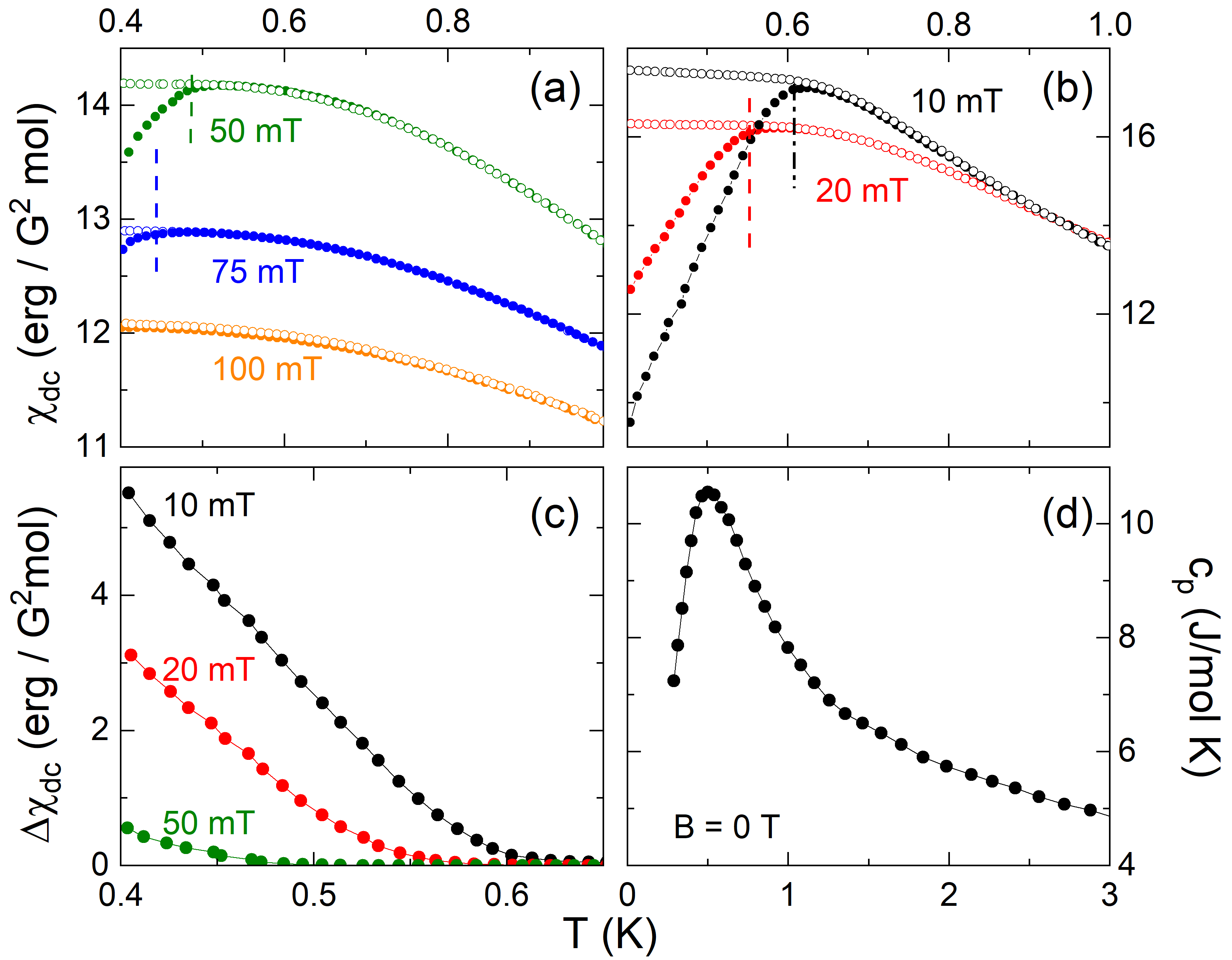}
    \caption{(a,b) Temperature dependence of the static magnetic susceptibility in various external magnetic fields $B \leq 0.1$~T. Empty and filled symbols refer to fc and zfc measurement protocols ($\chi_{\rm fc}$ and $\chi_{\rm zfc}$), respectively. Dashed lines indicate the blocking temperature $T_{\rm b}$. To correct for remanent field of the coil, the data have been normalised to the $M$ vs. $B$ data in Fig.~\ref{hysteresis}. (c) Difference between fc and zfc magnetic susceptibility $\Delta\chi = \chi_{\rm fc}-\chi_{\rm zfc}$ measured at different external fields. (d) Specific heat at $B=0$~T from Ref.~\cite{Sheetal_2022}.
}

    \label{lowT_chi}
\end{figure}

Figures~\ref{lowT_chi}a and b show the DC magnetic susceptibility of \HZO\ measured in the low-temperature regime between 400~mK and 1~K in external applied magnetic fields of 10~mT~$\leq B \leq$~100~mT. At $B = 10$~mT, the susceptibility shows a clear history dependence below $T\approx 0.6$~K as demonstrated by the difference between zfc and fc data. $\chi_{\rm zfc}$ exhibits a broad peak centered at $T\approx 0.6$~K and strongly decreases below this temperature but it does not reach zero. In contrast, $\chi_{\rm fc}$ is almost constant in the bifurcation regime which is a typical signature of a spin-frozen state~\cite{nagata1979low}. This behavior agrees well with other reports on pyrochlore systems where similar blocking temperatures $T_{\rm b} \approx 0.60 - 0.75$~K are observed for A$_2$B$_2$O$_7$ (A = Dy, Ho and B = Ti, Sn)~\cite{petrenko2011titanium,matsuhira2000low,snyder2004low,krey2012first}. The presence of similar blocking temperatures in systems with strongly different magnetic moments and hence strongly different magnetic couplings $J_{\rm nn}$ between the rare-earth moments in the pyrochlore structure indicates that \tb\ is rather independent on $J_{\rm nn}$~\cite{den2000dipolar}. The irreversibility between zfc and fc magnetic susceptibilities of \HZO\ persists under the applied magnetic fields and the blocking temperature shifts to lower values upon increasing the external field until no irreversibility is visible in the accessible temperature regime at $B\leq 0.1$~T (see Fig.~\ref{lowT_chi}a). At $B = 0$~T, the specific heat capacity shows a broad peak centered at $T = 0.5$~K as depicted in Fig.~\ref{lowT_chi}d. The nature of the peak excludes the presence of a magnetic phase transition in \HZO. Instead, it has been assigned to nuclear heat capacity resulting from the hyperfine interactions of \hoion\ nuclei magnetic moment with the effective magnetic field arising from the 4f electrons~\cite{Sheetal_2022}.

 \begin{figure}[htb]
    \centering
    \includegraphics[width=1\columnwidth,clip]{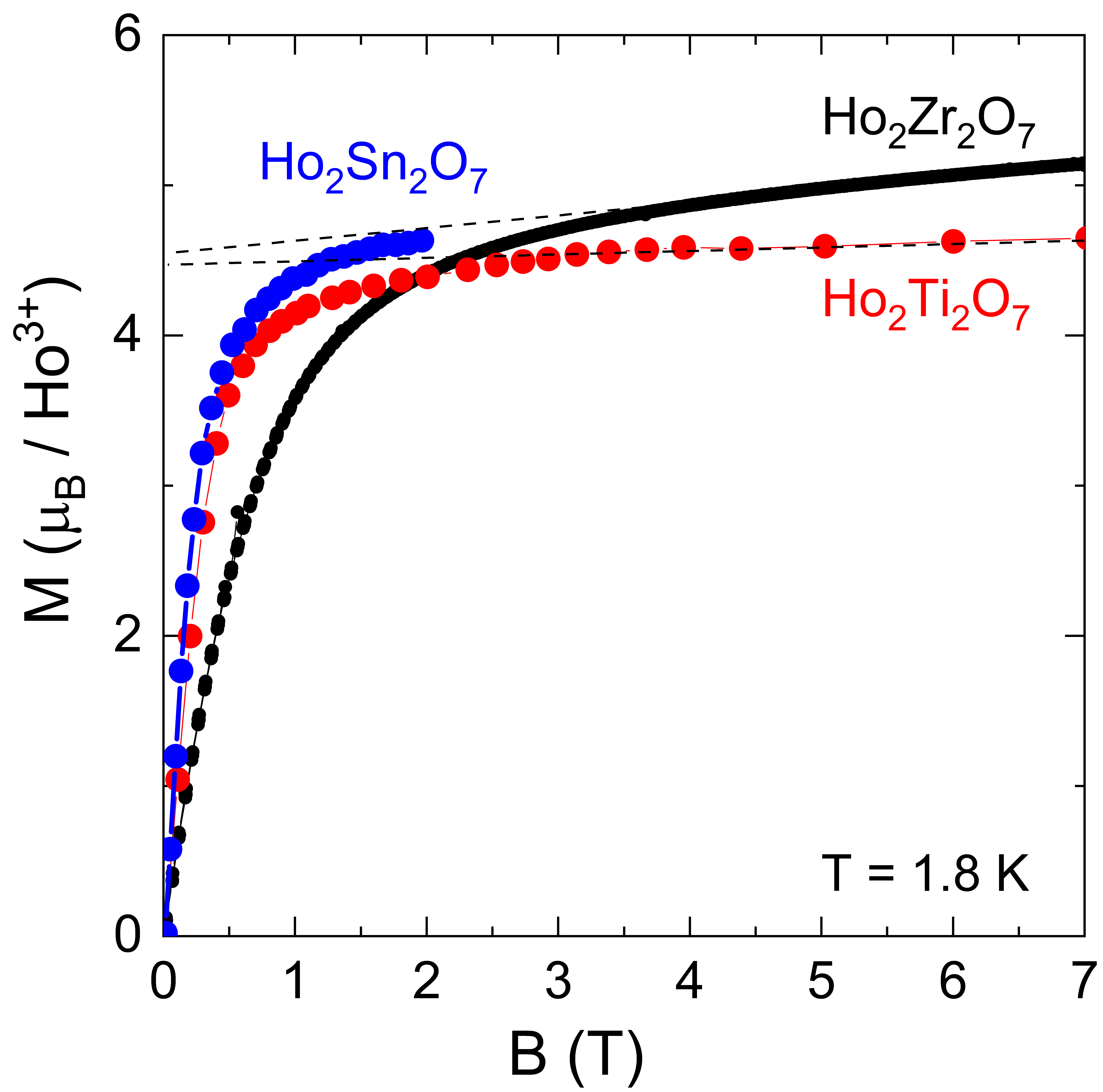}
    \caption{\Ahmed{Magnetic field dependence of the isothermal magnetization $M(B)$ measured in the upsweep mode at 1.8~K for \HZO, \HTO\ and \HSO\ polycrystalline samples. The data of \HTO\ and \HSO\ are taken from Refs.~\cite{bramwell2000bulk,matsuhira2000low}}. Dashed lines are guides to the eye.
}
    \label{MB}
\end{figure}

 \begin{figure}
    \centering
    \includegraphics[width=1\columnwidth,clip]{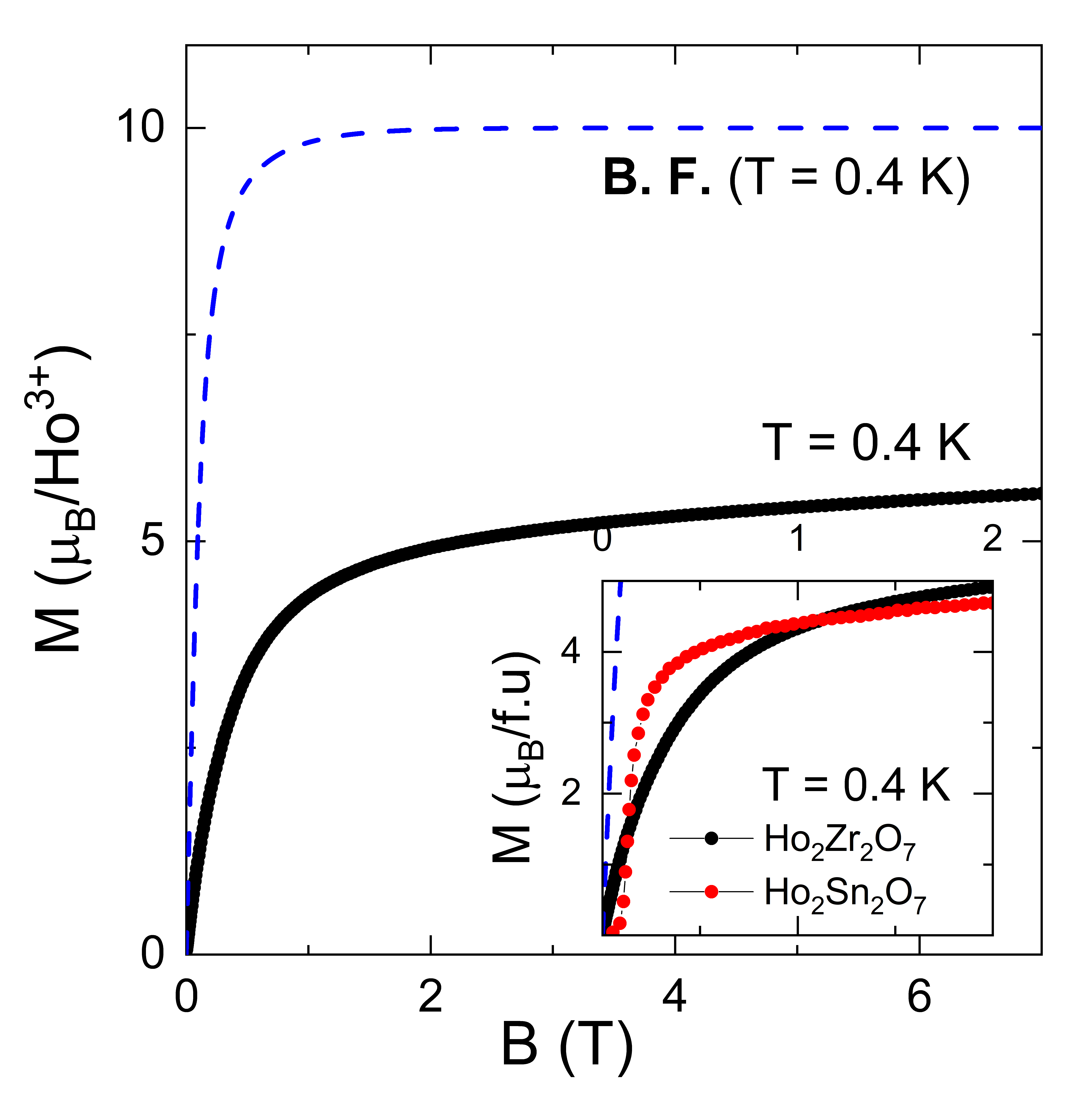}
    \caption{Magnetic field dependence of the isothermal magnetization of \HZO\ measured at $T = 0.4$~K in the upsweep mode and the dashed blue line represents the simulation of Brillouin function (B.F.) which corresponds to non-interacting \hoion\ magnetic ions with g$_{\mathrm J}$ = 1.25, J = 8 at $T = 0.4$~K. Inset: a comparison of the magnetizations of \HZO, and \HSO\ which is adopted from Ref.~\cite{matsuhira2000low}.
}
    \label{MB_Bril}
\end{figure}

At 1.8~K, the isothermal magnetization $M$ vs. $B$ of \HZO\ initially increases steeply and shows pronounced right-bending towards the saturation at higher fields (see Fig.~\ref{MB}). At $T=0.4$~K, $M$(7~T) amounts to 5.6~\mbho ; linearly extrapolating the high-field regime to $B=0$~T yields $\simeq 5.0$~\mbho\ (see Fig.~\ref{MB}). These values are much smaller than the theoretically expected saturation magnetization of $M_{\rm s} = 10$~\mb\  ($g = 1.25$, $J = 8$) of free \hoion\ ions as $M$(0.4~K,7~T) is only about 55\% of $M_{\mathrm s}$ (see Fig.~\ref{MB_Bril}). Such a reduced value can be attributed to substantial single ion anisotropy and it is consistent with the saturation magnetization value recently obtained for \HZO~\cite{ramon2020geometrically}, and for other pyrochlores such \HTO~\cite{krey2012first,bramwell2000bulk}, \HSO~\cite{matsuhira2000low}, \CZO~\cite{gao2019experimental}, and \PZO~\cite{kimura2013quantum}.
At small magnetic fields, the magnetization $M(B)$ of both \HZO\ and \HSO, measured at $T=0.4$~K, shows a left bending before the saturation evolves (see Fig.~\ref{MB}). This behaviour is reflected by corresponding peaks in the magnetic susceptibility $\partial M(B)/\partial B$ (see Fig. 1 in the SM~\cite{SM}) at $B = 0.03$~T (\HZO ) and $B = 0.13$~T (\HSO ). \Ahmed{In \HSO, $\partial M(B)/\partial B$ shows a very sharp peak, in contrast to the broader and smaller peak visible in \HZO. This illustrates the overall slower field-induced increase of $M(B)$ in \HZO. Further, the saturation in \HZO\ is much less rapidly achieved than expected for free moments (see the comparison with the Brillouin function in Fig.~\ref{MB_Bril}). The comparably modest increase of $M(B)$ in \HZO\ indicates the presence of antiferromagnetic short range correlations in the system, which agrees with the sign of the Weiss temperature estimated above. In contrast, the steeper increase of $M(B)$ in \HTO\ and \HSO\ (see Figs.~\ref{MB} and~\ref{MB_Bril}) is consistent with the presence of ferromagnetic correlations and exchange coupling reported in Refs.~\cite{bramwell2000bulk,matsuhira2000low}.}

\begin{figure}
    \centering
    \includegraphics[width=1\columnwidth,clip]{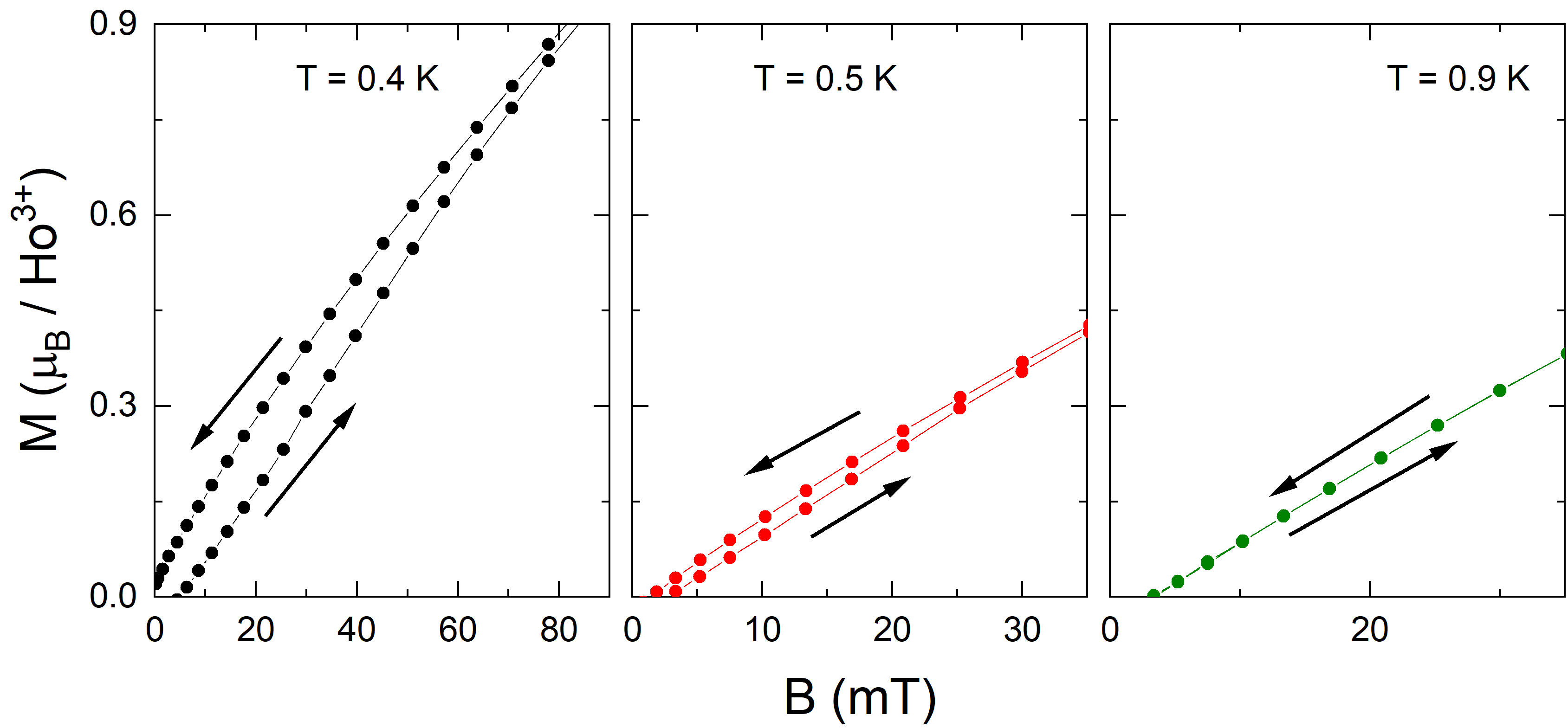}
    \caption{Magnetic field dependence (up and down sweeps) of the isothermal magnetization of \HZO\ measured at 0.4~K, 0.5~K and 0.9~K.
}
    \label{hysteresis}
\end{figure}

Spin freezing in \HZO\ which is evidenced by the irreversible behaviour of the DC magnetic susceptibility in Fig.~\ref{lowT_chi} also appears in the hysteresis of the isothermal magnetization measured below $T = 0.6$~K as depicted in Fig.~\ref{hysteresis}. The isothermal magnetization exhibits small and narrow loops at 0.4~K and 0.5~K with a width of 9~mT and 3~mT, respectively. In the $M(B)$ curves, we observe an initial left-bending in the up-sweep resulting in a broad hump in $\partial M/\partial B$. For instance, in $M(B)$ curve at T = 0.4~K, this broad hump is centered at around 300~mT; in the down-sweep, there is no such hump. As expected, the magnetization becomes reversible at higher temperatures as evidenced by the loop being fully closed at $T=0.9$~K. Note, despite the presence of clear hysteresis at $T < T_{\rm b}$, no conventional long-range thermodynamic ordering transition is observed as a function of temperature. This is evident by invoking the difference between $\chi_{\rm fc}$ and $\chi_{\rm zfc}$ measurements of the DC magnetic susceptibility, $\Delta\chi = \chi_{\rm fc} - \chi_{\rm zfc}$, as shown in Fig.~\ref{lowT_chi}c. Clearly, $\Delta\chi$ increases monotonically and smoothly through \tb\ with decreasing the temperature at all applied magnetic fields and there is no evidence of any anomalies around or below \tb.

\section{Discussion}

\begin{figure}[htb]
    \centering
    \includegraphics[width=1\columnwidth,clip]{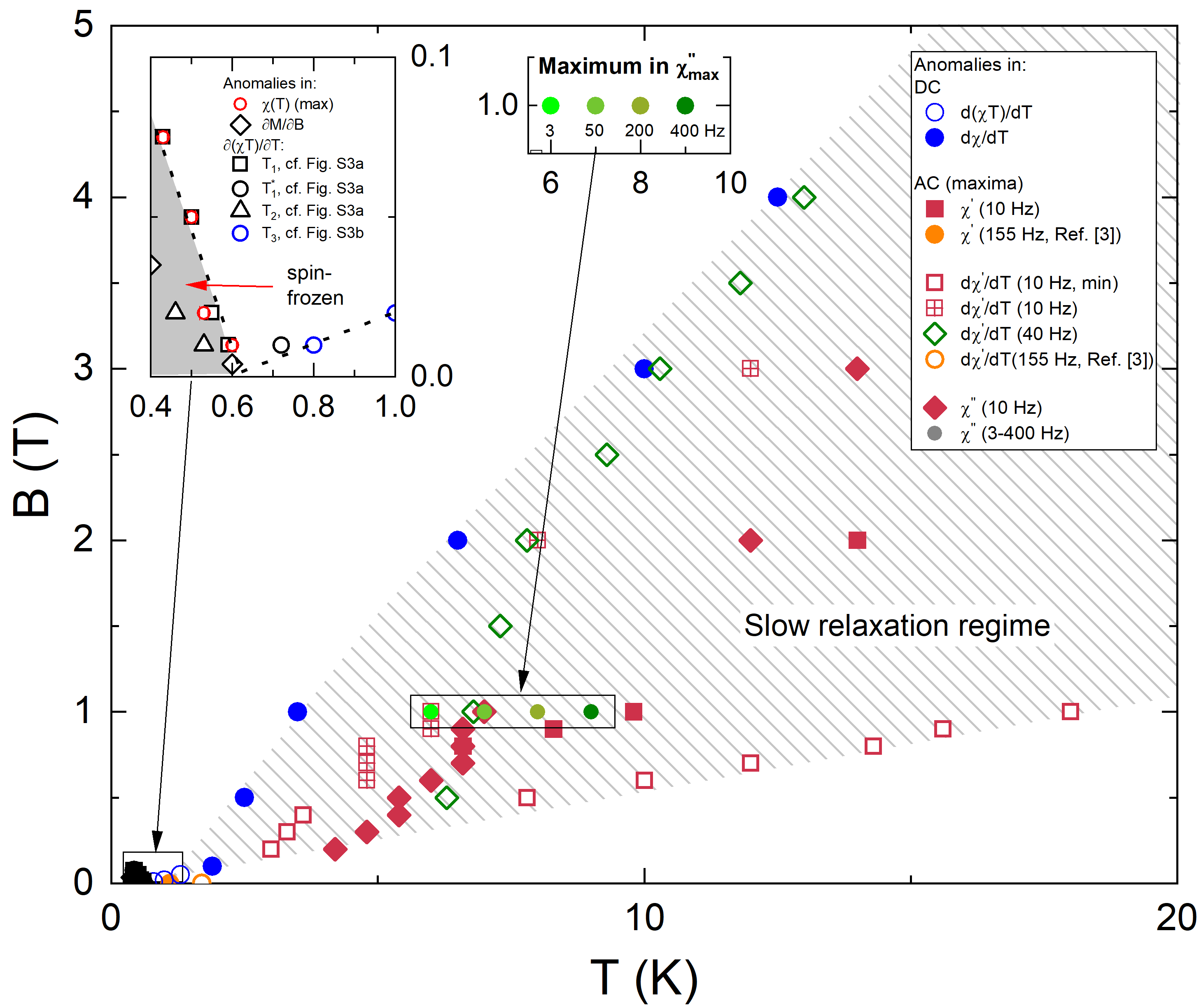}
    \caption{How spins freeze in \HZO . The data indicate anomaly features as obtained from Figs.~\ref{ac_infield}, \ref{derivative}, \ref{lowT_chi}, and from the SM. Below $T = 0.6$~K the system enters a spin frozen state. Insets highlight the low-temperature region as well as subsequent slowing down of spin dynamics as indicated by the maxima in $\chi''_{\rm ac}$ at frequencies 400 to 3~Hz. Error bars have been omitted for clarity.
}
    \label{phd}
\end{figure}

Our heat capacity, DC and AC magnetization studies show that \HZO\ does not show long-range magnetic order at least down to $T = 280$~mK. Instead, it has a disordered ground state with short-range antiferromagnetic correlations which evolves into a frozen spin state below $T_{\rm b} = 0.6$~K. While application of rather small magnetic fields ($B < 0.1$~T) suppresses $T_{\rm b}$, we observe field-induced features at higher temperatures. These additional features appear as maxima in $\chi'_{\rm ac}$ and $\chi''_{\rm ac}$ only in the presence of magnetic fields of $B > 0.1$~T (see Figs.~\ref{ac_infield} and~\ref{ac_arrhenius}). Both the DC and AC susceptibility data allow us to follow the magnetic field dependence of these several features as summarized in the phase diagram in Fig.~\ref{phd}. We also note that AC susceptibility in Ref.~\cite{ramon2020geometrically} measured as a function of temperature down to 50~mK and $B = 0$~T shows a maximum in $\chi'_{\rm ac}$ at around 1.1(1)~K. This reported maximum has been found to shift with frequency and obeys an Arrhenius-law with characteristic energy $E_{\rm b} \simeq 28$~K.

\begin{table*}
    \caption{Comparison of the relaxation processes and their related parameters in $R_2B_2$O$_7$; $R$ = Ho, and Dy, $B$ = Ti and Zr. $B_{\rm dc}$, $T_{\mathrm g}$, and $\tau_0$ stands for applied magnetic field, freezing temperature, and characteristic time, respectively.}
    \vspace{3mm} % Adjust the height of the space between caption and tabular
    \centering
    \begin{tabular}{c c c c c c c}
\hline
 Material & $B_{\rm dc}$ (T) & $T_{\mathrm g2}$ (K) & $T_{\mathrm g1}$ (K) & $\tau _o^2$ (sec) & $\tau _0^1$ (sec) & Ref. \\
 \hline\hline
 \DTO & 0 & 1 & 15(1) & $5 \times 10^{-11}$ & $2.2 \times 10^{-10}$ &~\cite{matsuhira2001novel,snyder2001spin} \\
 \HTO & 0 & 1 & - & $1.6 \times 10^{-11}$ & - & ~\cite{ehlers2004evidence,ehlers2002dynamical} \\
 \HZO & 0 & 1 & - &  $8 \times 10^{-13}$ & - &~\cite{ramon2020geometrically} \\
 \HTO & 1 & - & 15(1) & - & $5 \times 10^{-8}$ &~\cite{ehlers2004evidence} \\
 \HZO & 1 & - & 7(1) &  - & $2.1 \times 10^{-8}$  &~This work \\
 \hline
\end{tabular}
\label{compare}
\end{table*}

Field-induced spin freezing anomalies at high temperatures far above the frozen spin state ($T_{\mathrm g1}(1~{\rm T})= 7(1)$~K~$\gg T_{\rm b} = 0.6$~K found at hand for \HZO ) seems to be an universal phenomenon in pyrochlore lattices with quasi-classical (large) magnetic moments. Specifically, the anomaly observed in \HZO's AC magnetization data at $T_{\mathrm {g1}} = 7(1)$~K and $B=1$~T resembles the peaks observed in the canonical spin ice systems \DTO\ and \HTO~\cite{snyder2001spin,ehlers2004evidence}, at $T = 15(1)$~K. This anomaly is also very similar to the peaks observed in $\chi'$ of the disordered-fluorite \TZO~\cite{ramon2023glassy} at $T = 25$~K and the pyrochlore system \TTO\ at $T = 20$~K~\cite{ueland2006slow}. To follow this phenomenon, the spin freezing temperatures of \DTO\ and \HTO\ observed at $T_{\mathrm {g1}} = 15$~K, $T_{\mathrm {g2}} = 1$~K and their relaxation characteristic times are listed in Table~\ref{compare} and compared with freezing temperatures of \HZO\ inferred from the current work as depicted in Fig.~\ref{ac_arrhenius} and from Ref.~\cite{ramon2020geometrically} at $T_{\mathrm {g2}} = 1$~K. At $B = 0$~T, the characteristic time reported for \HZO~\cite{ramon2020geometrically} from the frequency dependence of $T_{\mathrm g2} = 1$~K is two orders of magnitude lower than the one reported for the canonical spin ice \HTO~\cite{ehlers2002dynamical,ehlers2004evidence}. While, at $B = 1$~T, the characteristic time ($\tau = 2.1 \times 10^{-8}$) estimated for \HZO\ from the frequency dependence of $T_{\mathrm g1} = 7(1)$~K (see Fig.~\ref{ac_arrhenius}) has the same order of magnitude as the one reported for \HTO.

Further information on the nature of magnetism in \HZO\ is obtained from the isothermal magnetisation data in Fig.~\ref{MB}. For a spin ice model with nearest-neighbor interactions, Monte Carlo simulations predict a magnetization plateau at 3.33~\mb/ion and a saturated moment of 5.00~\mb/ion for the magnetization measured along $B ||$ [111] of the pyrochlore single crystal~\cite{harris1998liquid}. This has been verified by experimental studies, at 1.8~K, on \HTO\ single crystals~\cite{cornelius2001short}. Since the system at hand is a polycrystal, we compare $M(B)$ $T = 1.8~{\rm K}$ with the magnetization of a \HTO\ polycrystal~\cite{bramwell2000bulk} as displayed in Fig.~\ref{MB}. This comparison yields the following conclusions: (1) The saturation moment, at $B = 7$~T, of \HZO\ amounts to 5.1 (1)~\mb/\hoion, exceeding the saturation magnetisation of \HTO\ (4.6~\mb/\hoion ). While our $M_{\rm s}$ agrees well with the spin ice model predictions, the magnetization of \HZO\ and \HTO\ polycrystals do not show the predicted plateaus; for \HZO, this is confirmed at low temperatures down to  $T = 400$~mK as shown in Fig.~\ref{MB_Bril}. (2) The magnetisation field-dependence in both systems is different, as the saturation sets in at lower magnetic fields in case of \HTO. The characteristic field scale in \HTO\ of $B = 0.6 (1)$~T is shifted to $B = 1.4 (2)$~T in \HZO. As discussed above, we conclude the presence of antiferromagnetic correlations in \HZO, in contrast to the ferromagnetic correlations found in \HTO~\cite{bramwell2000bulk,cornelius2001short}.

Qualitatively, the irreversible DC magnetic susceptibility, and the absence of a magnetic phase transition down to 280~mK, combined with the frequency dependence of the maximum visible in $\chi'_{\rm ac}$ at 1~K~\cite{ramon2020geometrically} suggest a spin freezing state in \HZO. These features are in a good agreement with spin-freezing observed in chemically disordered systems as well as in site-ordered geometrically frustrated antiferromagnets~\cite{ramirez2001handbook,ramirez1990strong,schiffer1997two,wills2001long,snyder2004low}. Further, with a closer look on the evolution of the position of the blocking temperature $T_{\rm b}$ with the magnetic field, we find that it is shifted to lower values and suppressed very fast already by low magnetic fields. Such magnetic field effect agrees very well with the findings in the spin ice system \HTO~\cite{krey2012first} but contradicts to what had been observed in the spin ice system \DTO~\cite{snyder2004low}. In the latter, $T_b$ not only persists at higher magnetic fields but also increases with increasing the applied magnetic field. Our data hence show that the applied magnetic field slows down spin dynamics in \HZO\ and promotes greater thermal stability of the regime below $T_b$ against thermal fluctuations.

Finally, the features appearing in $\chi_{dc}$, $M(B)$, \chip, \chipp, and their derivatives are used to construct the phase diagram of \HZO\ displayed in Fig.~\ref{phd}. This phase diagram comprises two separate regions corresponding to a high-temperature slow spin relaxation regime and a spin-frozen regime at low temperatures. The slow spin relaxation region (dashed area) is bordered by the anomalies visible in the derivatives of DC and AC magnetic susceptibilities. Anomalies in this region appear in finite DC magnetic fields and the region spans a wide temperature regime from $T = 18(2)$~K down to $T = 1.1(1)$~K. The fact that anomaly positions not only depend on temperature and magnetic field but also on the frequency implies the successive freezing of spin upon cooling. Specifically, at a given field the characteristic temperatures increase with the frequency in AC measurements. On further cooling, \HZO\ enters a frozen state below $T_{b} = 0.6$~K. This phase resembles the frozen phase reported in \HTO\ single crystals as reported in Ref.~\cite{krey2012first} where it has been also derived from magnetization data (measured along $B ||$ [111] axis) and also appears below 0.6~K. Notably, pyrochlore structures with the formula A$_2$B$_2$O$_7$ (A = Dy, Ho and B = Ti, Sn) exhibit $T_{\rm b} \approx 0.60 - 0.75$~K~\cite{petrenko2011titanium,matsuhira2000low,snyder2004low,krey2012first}, i.e., $T_{\rm b}$ is rather independent on $J_{\rm nn}$~\cite{den2000dipolar}.

\section{Conclusions}

In conclusion, our heat capacity, DC, and AC magnetization studies show that \HZO\ does not evolve a long-range magnetic order down to $T = 280$~mK. Instead, it has a disordered magnetic ground state with short-range antiferromagnetic correlations. Further, due to geometrical frustration, the spins tend to form a spin frozen state below $T = 0.6$~K. \Ahmed{Our results imply that the blocking temperatures $T_{\rm b} \approx 0.60 - 0.75$~K observed in A$_2$B$_2$O$_7$ (A = Dy, Ho and B = Ti, and Sn) are barely dependent on the magnetic couplings $J_{\rm nn}$ between the rare-earth magnetic ions in the ordered fluorite (pyrochlore) or in the disordered-fluorite structure in case of \HZO. Finally, the field-induced spin freezing anomalies at high temperatures far above the frozen spin state seems to be a universal phenomenon in the ordered fluorite (pyrochlore) as well as disordered-fluorite lattices with quasi-classical (large) magnetic moments.}

\section*{Supplemental material}

\begin{figure}[h]
    \centering
    \includegraphics[width=1\columnwidth,clip]{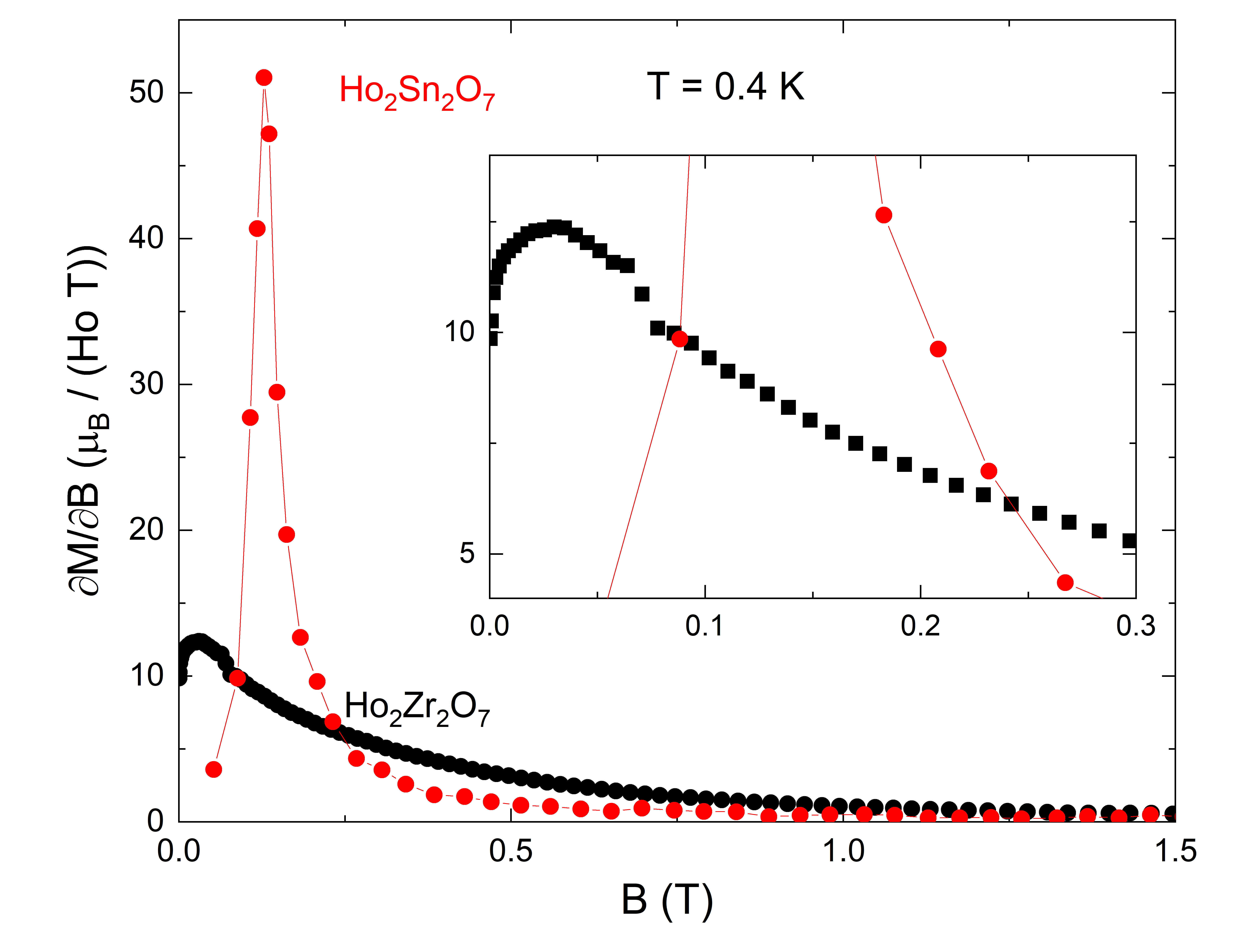}
    \caption{Derivatives of the isothermal Magnetisation M(B) of \HZO, and \HSO\ measured at $T = 0.4$~K. The data of \HSO\ are adopted from Ref.[27]. Inset: A narrow scale of the main plot to emphasize the anomalies appear at $B < 0.3$~T.}
    \label{dMdB}
\end{figure}

\begin{figure}[h]
    \centering
    \includegraphics[width=1\columnwidth,clip]{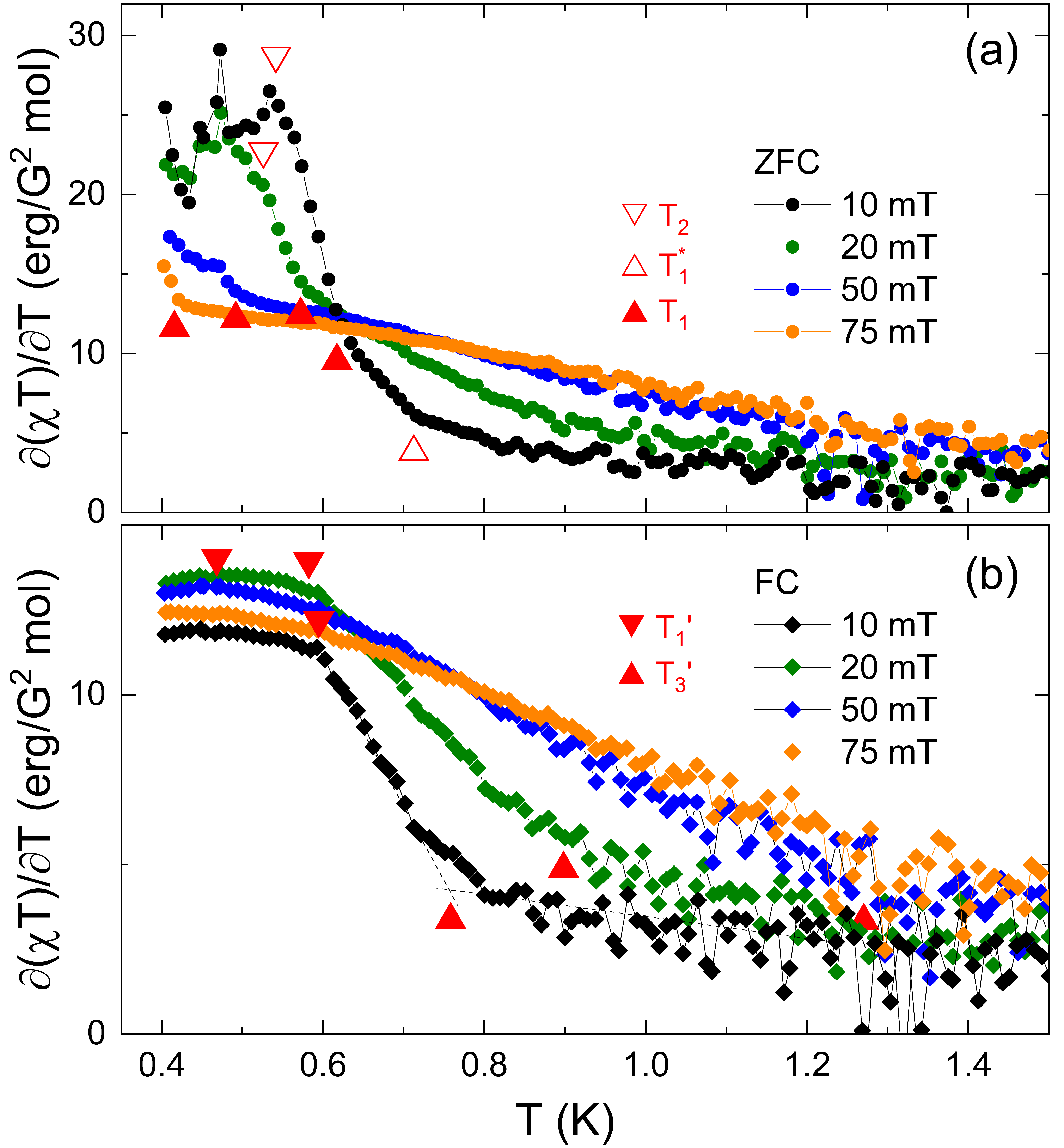} %scale=0.5
    \caption{Temperature dependence of the magnetic heat capacity of \HZO\ derived from the DC magnetic susceptibility which is measured at $B \leq 75$~mT: (a) in zfc mode and (b) fc mode. Triangles mark the features appeared in the data.}
    \label{d(XT)dT_lowT}
\end{figure}

%\vspace{5cm}

\section{Acknowledgments}

We thank Nang Tan for valuable discussions. We acknowledge financial support by Deutsche Forschungsgemeinschaft (DFG) under Germany’s Excellence Strategy EXC2181/1-390900948 (the Heidelberg STRUCTURES Excellence Cluster) and through project KL 1824/13-1. A.E.~acknowledges support by DAAD through the GSSP program.

\vspace{5cm}

\bibliography{Refs}
\end{document}